\documentclass[10pt,a4paper]{iopart}

\usepackage{graphicx}
\usepackage{hyperref}
\usepackage{iopams}

\begin{document}
\title{Guided-mode resonance filter extended-cavity diode laser}

\author{Lauren Guillemot$^{1,2}$, Thomas Oksenhendler$^{1}$, Sylvain Pelloquin,$^{3}$,Olivier Gauthier-Lafaye$^{3}$, Antoine Monmayrant$^{3}$, Thierry Chaneli\`ere$^{4,5}$}

\address{$^{1}$ iTEOX, 14 Avenue Jean Jaur\`es, 91940 Gometz-le-ch\^atel, France}
\address{$^{2}$Normandie Univ, ENSICAEN, UNICAEN, CEA, CNRS, CIMAP, 14000 Caen, France}
\address{$^{3}$LAAS-CNRS, Universit\'e de Toulouse, CNRS, 31400 Toulouse, France}
\address{$^{4}$Laboratoire Aim\'e Cotton, CNRS, Univ. Paris-Sud, ENS-Cachan, Universit\'e Paris-Saclay, 91405, Orsay, France}
\address{$^{5}$Univ. Grenoble Alpes, CNRS, Grenoble INP, Institut N\'eel, 38000 Grenoble, France}

\ead{thierry.chaneliere@neel.cnrs.fr}
\begin{abstract}
We characterize a semiconductor external cavity diode laser whose optical feedback is provided by a guided mode resonance filter (GMRF). We focus on the spectral properties.
The wavelength of operation falls in the telecom range (1506\,nm).
The GMRF acting both as a wavelength intracavity filter and feedback mirror allows for a compact laser design.
The single-mode operation is verified in a wide range of driving currents. We finely tune the cavity length to adjust the frequency by 14\,GHz without mode-hops in agreement with the expected free-spectral range of the resonator $\sim$20\,GHz.
The compactness of the cavity allows fast frequency sweeps when modulating the current (90\,MHz/mA at 100\,kHz, the modulation bandwidth).
The frequency noise (366\,kHz white-noise contribution) is also analysed to evaluate the potential of our design for high-resolution applications.
\end{abstract}


%\submitto{\LP}
\maketitle
\ioptwocol

\section{Introduction}

%Diode lasers are extensively used for their compactness, tunability and low cost. However, some applications such as atomic physics \cite{chu1991laser} or spectroscopy \cite{sowoidnich2010application}, %mettre un truc plus prÃ©cis, chercher d'autre ref
 %require high spectral purity and diodes sources are often multimode and exhibit poor frequency stability. 
 
External cavity diode lasers (ECDLs) are a standard way of building single-mode and narrow-linewidth sources \cite{petermann1995external}, while keeping the intrinsic benefits of the semiconductor lasers such as compactness, low cost and high-modulation capability in amplitude and frequency.    
Traditional ECDLs design uses a bulk diffraction grating in Littrow \cite{hawthorn2001littrow} or Littman/Metcalf \cite{littman1978spectrally} configurations leading to high sensitivity to vibrations and temperature drift, and limited frequency agility.
In this lineage, the recent developments in fabricating volume holographic grating allowed to conceive more compact cavities \cite{luvsandamdin2014micro,luvsandamdin2013development,paboeuf2008narrow,rauch2015compact,repasky2003tunable,scheuer2005lasing,hieta2009external,shen2011mode} even if the bulk character of the frequency filter prevents a rapid cavity tuning other than the diode driving current.

%In order to overtake these issues, several approaches has been proposed. Distributed Bragg Reflectors (DBR) \cite{sumpf2014wavelength} with a spectral selective mirror at one side of the cavity; interference filter \cite{baillard2006interference} \cite{Zorabedian:88} in a cateye configuration within an external cavity formed by a mirror. %Volume Holographic Gratings (VHGs) \cite{luvsandamdin2014micro} \cite{luvsandamdin2013development}.

In the quest of smaller footprint ECDLs, several approaches have been proposed.
The first one makes use of an interference filter as selective element. The latter fully benefits from thin-film technologies and nanolithography techniques \cite{Zorabedian:88} with a interesting potential in terms of mass production. Low vibration sensitivity has been obtained in the so-called cateye configuration for the external feedback mirror \cite{baillard2006interference}.
Still, the use of different optical elements for feedback and frequency selection limits the size and the versatility of the system.

More recent approaches consists in the direct coupling of the semiconductor laser into an integrated resonant structure as a whispering gallery mode resonator \cite{liang2015ultralow}, a nano-fabricated ring resonator \cite{lin2018characterization} or a Bragg reflector on a planar silica waveguide  \cite{stolpner2008low,numata2010performance}.
These hybrid architectures have been pushed to a remarkable level of integration by producing butterfly-packaged ECDL \cite{assembly}.
They also provide ultralow-noise feature as compared to distributed feedback \cite{dougherty1999semiconductor} and distributed Bragg reflector \cite{sumpf2014wavelength} on-chip design.

Fully integrated photonic circuits have their own advantages when narrow-linewidth are targeted. Initially designed for a wide-tunability by carrier injection in sampled distributed Bragg reflector (DBR), the tuning mechanism induces extra-losses that limit the linewidth \cite{coldren2012diode}. Thermal tuning (without carrier injection) appears as a possibility at the price of a slower operation and higher thermal heating \cite{Duncan}.  Heterogeneous integration of III-V materials on silicon allows to fully integrate the gain chip (III-V semiconductor) and the resonator in low-loss silicon circuitry \cite{Bowers, Oldenbeuving_2012}.

These multiplicity of material platforms and architecture somehow reveals the diversity of applications. The latter cover different domains as atomic and molecular spectroscopy for which technology transfer from lab research requires a minimum level of integration while maintaining a narrow-linewidth \cite{svanberg2012atomic}. The development of advanced digital modulation schemes in modern telecommunications combining amplitude, phase and/or frequency modulations has also shown the need for narrowband lasers to fully exploit the potential of large constellation size \cite{seimetz2008laser}.

In this paper, we report on an alternative approach based on a Guided Mode Resonance Filter (GMRF) as feedback element for the ECDL.
A GMRF is a reflecting narrow-band filter composed of a subwavelength grating on a waveguide \cite{wang1993theory, rosenblatt1997resonant}.
Such devices can exhibit a high reflection approaching 100\,\% \cite{liu1998high} and a FWHM (Full Width at Half Maximum) resonance as narrow as 0.12\,nm \cite{levy2000very} making them competitive as compared to volume Bragg gratings in terms of spectral selectivity. The angle tuning of volume gratings or interference filters is not directly possible with a single period GMRF but can be obtained with a chirped grating (translation tuning) without putting hard constraints on the fabrication technique.

A GMRF is a planar reflecting optical element benefiting from nano fabrication techniques with a broad range of potential applications in sensing, solar energy, photodetection, polarizing optics or spectrometry \cite{quaranta2018recent}.
Similar to the previously mentioned  hybrid architectures, GMRF-based lasers can be packaged to small footprint \cite{kondo2015design}.
In a sense, GMRFs are at the interface between volume gratings and edge-coupled wave-guided structures (gallery mode, ring or planar Bragg resonators). As volume gratings, they are used in retro-reflection (and not edge-coupled). As planar elements, they contain a wave-guiding resonant structure.

There have been a number of studies investigating the potential of GMRFs for laser frequency stabilization, first on dye lasers \cite{avrutskii1985light,avrutskiui1986spectral,kobayashi2005surface} and large-area or disk semiconductors lasers \cite{avrutsky2001waveguide, Giet:07}, then on fibered  \cite{sims2009narrow,mehta2007guided,li2012guided,sims2011spectral} or solid-state lasers \cite{Vogel:12, Aubourg:14}.
Despite a clear potential, only few ECDLs using GMRF as external mirror have been demonstrated.
The first implementation also demonstrates an interesting 1\,nm thermal tuning range \cite{block2005semiconductor}.
The latter can be pushed to 7\,nm by using a liquid-crystal GMRF \cite{chang2007tunable}.
Blue laser diode have also been stabilized by a variable-reflectivity GMRF \cite{byrd2014blue}.
Poor angular tolerance inherent to GMRF can be overcome by a sophisticated design of the in-plane resonant structure in the so-called cavity resonator integrated guided mode filter configuration \cite{buet2012wavelength}.

In this letter, we present an ECDL operating at 1.5\,$\mu$m wavelength using a GMRF as external mirror. Our prototype is not integrated but for our demonstration we aim at maintaining a short cavity (few mm).
After detailing the extended cavity design, we show that the ECDL is essentially single-mode over a wide range of driving currents.
We also evaluate the frequency modulation response when the diode current and the cavity length are varied.
We finally characterize the white-noise components of the frequency noise (quantum-limited linewidth). The latter should scale inversely as the extended cavity length so justifying the ECDL setup as compared to a single-ridge laser diode.

\section{Laser design}

\begin{figure}[htp]
\centering
\includegraphics[width=.9 \columnwidth]{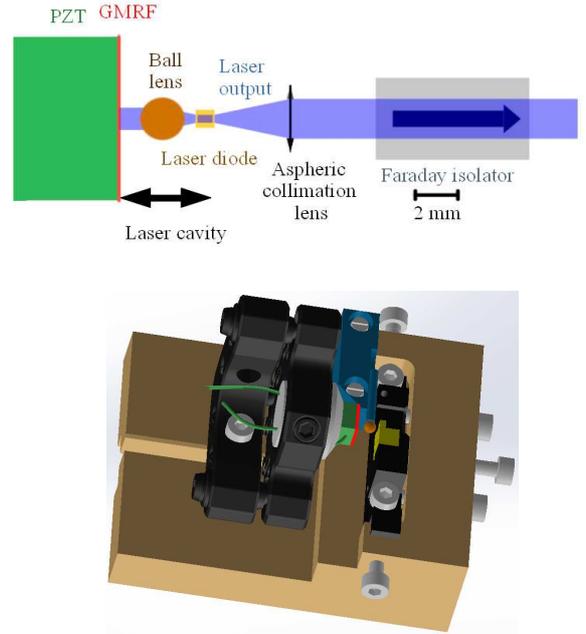}
\caption{Top:Extended-cavity schematic.
The cavity is composed of the diode, a collimation ball lens and the GMRF.
The output from the back facet of the diode is collimated (aspheric lens) and injected into a single mode fiber after a miniature Faraday isolator to avoid unwanted optical feedback into the diode. Bottom: Extended-cavity design.
The gain chip is an AR coated diode (yellow).
A 2-mm ball lens (orange) collimates the beam.
The GMRF acts as filtering element and end mirror (red).
A PZT is used to adjust precisely the cavity length (green).
Adjustment screws (push-pull configuration) are used to align the diode and the ball lens.
We align the end mirror of the cavity with 1/2" mirror mount (Thorlabs KM05 in black)}
\label{fig:scheme_ensemble}
\end{figure}

We choose for our beam geometry a generic tunable ECDL configuration \cite{zorabedian1995tunable} as used in the early days for interference-filter based cavity design \cite{Zorabedian:88}.
Our gain medium is Fabry-Perot single stripe diode with anti-reflection (AR) coating on one side (see fig.\ref{fig:scheme_ensemble},top).
The uncoated back facet (reflectivity $\sim$30\,\%) serves as an output mirror of the external cavity.
The AR coated facet reflectivity can be estimated from the amplified spontaneous emission spectrum of the diode.
This latter is modulated by the Fabry-Perot effect in the gain stripe.
The modulation contrast gives an estimated AR coated facet reflectivity of 0.7\,\% at our wavelength of operation (1506\,nm).
Inside the cavity, we use a 2-mm ball lens to collimate the beam.
The GMRF serves both as a filtering element and as an end mirror.

%\begin{figure}[htp]
%\centering
%%\includegraphics[width=0.8\textwidth]{vue2.png}
%\includegraphics[width=0.5\textwidth]{vue2_cropped.png}
%\caption{Extended-cavity design.
%The gain chip is an AR coated diode (yellow).
%A 2-mm ball lens (orange) collimates the beam.
%The GMRF acts as filtering element and end mirror (red).
%A PZT is used to adjust precisely the cavity length (green).
%Adjustment screws (push-pull configuration) are used to align the diode and the ball lens.
%We align the end mirror of the cavity with 1/2" mirror mount (Thorlabs KM05 in black).}
%\label{fig:design}
%\end{figure}

\begin{figure}[htp]
\centering
\includegraphics[width=0.8\columnwidth]{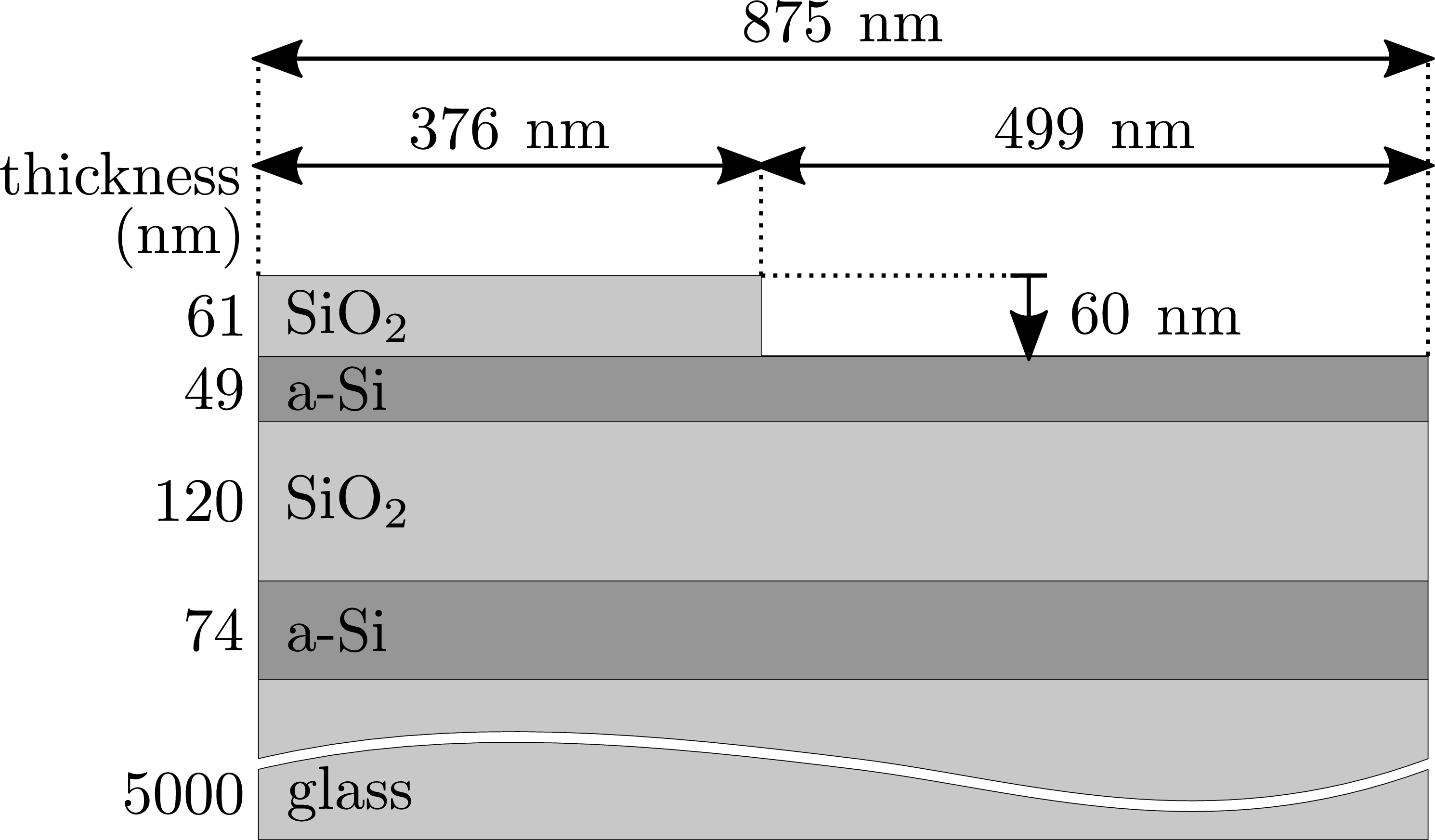}
\caption{GMRF: one period of the GMRF structure used as a reflector.}
\label{fig:gmrf}
\end{figure}

The geometry of the GMRF is depicted in fig.\ref{fig:gmrf} (only one period is represented). It consists of an SiO$_2$/air grating (875\,nm period with 60\,nm-deep 376\,nm Si0$_2$ ridges and 499\,nm air grooves) ontop of a a-Si/SiO$_2$ multilayer on a glass substrate. Out of resonance, around 1506\,nm, the multilayer acts simultaneously as an antireflection coating at normal incidence and a single-mode planar waveguide.
At resonance, the grating couples incident waves to the planar guided mode resulting in a 2\,nm-large reflective peak at 1506\,nm.

The GMRF is fabricated by soft-mold nano-imprint lithography, as reported in \cite{pelloquin2017soft} on a 4-inch wafer. First, a multilayer stack is deposited by  inductively-coupled-plasma plasma-enhanced chemical vapor deposition (ICP-PECVD) on a glass substrate. This multilayer acts both as an AR coating and a waveguide at the wavelength of design.
Second, the grating is etched in the first layer of the stack. A master grating is defined by e-beam lithography and etched in a silicon wafer. Using thermal nano-imprint lithography (T-NIL), this master grating is replicated onto a soft mold. This soft mold is then used in a UV nano-imprint lithography step to transfer the grating pattern in a layer of resist deposited ontop of the final sample. A last etching step is used to transfer the grating from the resist into the first layer of the multistack. Eventually, the wafer is diced into 5x5\,mm$^2$ individual filters.
Under normal incidence, the GMRF exhibits a $85\,\%$ reflectivity peak at $1506\,$nm with a FWHM of 1.8$\,$nm (see \cite{pelloquin2017soft} for a detailed description of the fabrication process and the final structure).

The GMRF is glued on a piezoelectric actuator PZT (Thorlabs TA0505D024W).
Because of the low footprint of the GMRF, the cavity can be really small (few millimetres in our case, see fig.~\ref{fig:scheme_ensemble}, bottom).
This allows an enhanced frequency tunability that we evaluate in this report. 
Additionally, the small mass of the GMRF, as compared to the popular grating based ECDL \cite{arnold1998simple}, permits faster operation of the quasi-unloaded PZT thus significantly increasing the agility of the laser.

The uncoated back facet of the diode is the output port of the extended-cavity.
The collected output power is typically 5\,mW at 100\,mA driving current\footnote{The current controller is a ILX Lightwave LCM-39427, combining current and thermoelectric cooling controller.}.
We obtain very comparable values than the previously demonstrated grating based ECDL with the same gain chip \cite{crozatier2006}.
After proper collimation of the output with an aspheric lens, we insert an optical isolator in the path to avoid unwanted feedback from the measurement setup. See fig.\ref{fig:scheme_ensemble}(top) for as schematic and fig.\ref{fig:scheme_ensemble}(bottom) for the corresponding design.
A fraction of mW is finally injected into a single mode fiber for the different characterizations reported hereafter.

\section{Continuous-wave operation and characterization}\label{cw}
In this section, we characterize the laser in continuous-wave (CW) operation when the two tuning parameters, namely the diode injection current and PZT voltage, are slowly varied.

\begin{figure}[htp]
\centering
\includegraphics[width=.9\columnwidth]{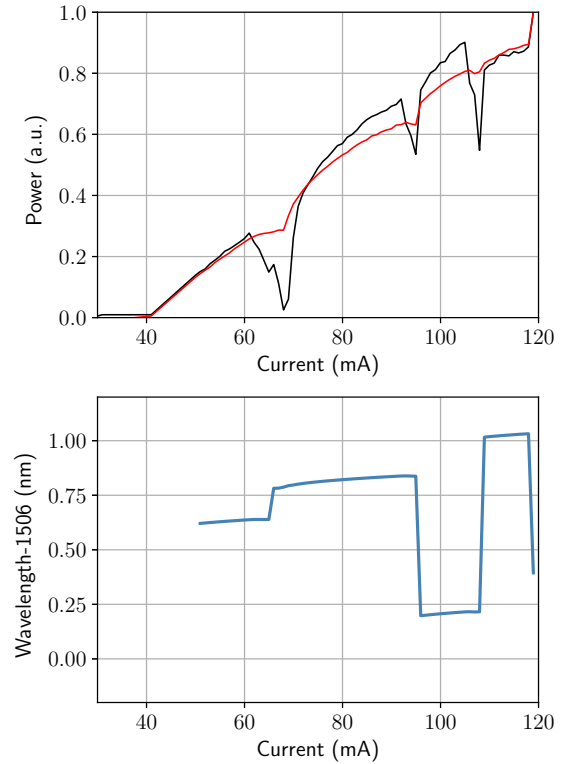}
\caption{Power (top) and wavelength (bottom) as a function of the diode injection current.
The power is monitored with a power-meter (red curve) but also by evaluating the mode amplitude (black curve) of a scanning Fabry-Perot interferometer in order to identify the mode-hops (see text for details).}
\label{fig:Trait_series_wavelength_FP}
%/home/thierry/Exp/LAC/alcor_TC/20180418_alcor_lambda_vs_I/
\end{figure}

We first record the output power and the wavelength with a HighFinesse/\AA ngstrom WS7-30 wavelength meter as function of the injection current (see fig.\ref{fig:Trait_series_wavelength_FP}).
The threshold current is typically 40\,mA and the maximum operation current 120\,mA.

\begin{figure}[htp]
\centering
\includegraphics[width=.9\columnwidth]{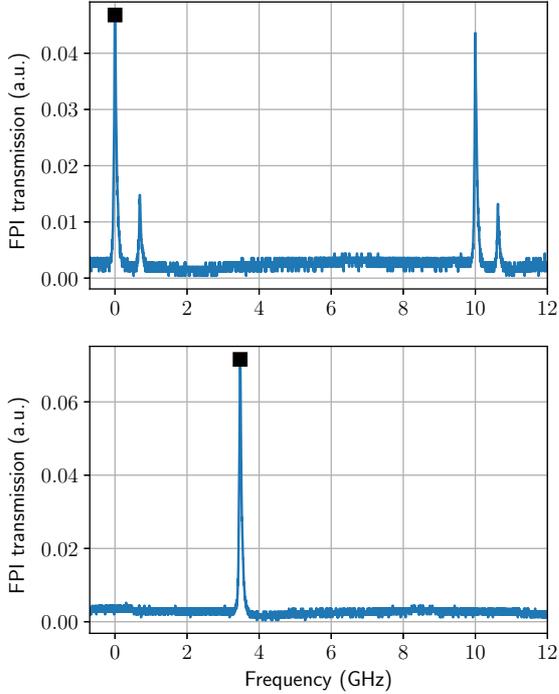}
\caption{Two examples of Fabry-Perot interferometer (FPI) spectra: at a mode jump around 95\,mA (top) and inside a single-mode region around 100\,mA (bottom).
The free spectral range of the scanning FPI is 10 GHz. The black squares indicate the amplitude of the highest mode, defining the mode amplitude as displayed in fig.\ref{fig:Trait_series_wavelength_FP} (top).}
\label{fig:Trait_deux_FP_article} 
%/home/thierry/Exp/LAC/alcor_TC/20180418_alcor_lambda_vs_I/
\end{figure}

The comparison between the output power and the mode amplitude from a scanning Fabry-Perot interferometer (FPI) Melles Griot 13SAF048as a function of the current indicates the presence of mode-hops at 65\,mA, 95\,mA, 108\,mA 118\,mA where power drops are visible (see fig.\ref{fig:Trait_series_wavelength_FP}, top and \ref{fig:Trait_deux_FP_article}). The displayed mode amplitude is defined from the scanning FPI measurements as illustrated in fig.\ref{fig:Trait_deux_FP_article}. This is the amplitude of the highest mode. The mode amplitude drops when a second mode appears (close to a mode-hop).

The wavelength measurements confirm a well established longitudinal single-mode operation above 50\,mA (below the measurements are inconsistent and are not displayed in fig.\ref{fig:Trait_series_wavelength_FP},bottom) and exhibit the mode-hops at 65\,mA, 95\,mA, 108\,mA 118\,mA  (see fig.\ref{fig:Trait_series_wavelength_FP}, bottom).
It should be noted that the mode drops of the FPI measurements are more reliable than the wavelength meter to evaluate the single-mode character. In the case of two competing modes, the wavelength meter gives the highest amplitude mode wavelength.

The mode jumps can be explained by the competition between the numerous extended-cavity modes which are contained within the FWHM of the GMRF (1.8$\,$nm $\approx$ 237\,GHz). When the current is adjusted, the optical length of the extended cavity is modified (diode index modulation). This sweeps the cavity modes and induces a mode jump. Indeed the smallest mode jump by $\sim$ 20\,GHz ($\sim$ 0.15\,nm at 65\,mA) gives an estimation of the extended-cavity free spectral range (FSR) corresponding to a cavity optical length of 7.5\,mm.
This is consistent with our cavity design composed of the laser diode (FSR of 90\,GHz corresponding to a 1.6\,mm optical length), the 2\,mm ball lens (focal length of 0.5\,mm and 3\,mm of optical propagation in the N-BK7 glass) and 2-2.5\,mm of free space propagation. So there are approximately a dozen of extended cavity modes contained in the FWHM of the GMRF.
%FSR diode = 90GHz -> 1.6 mm
% EFL ball lens = n*D/4/(n-1), D=2, n=1.52 -> EFL=1.4615 soit Focal Length of Ball Lens = .5mm

To further characterize the single-mode operation, we monitor the transmission of the scanning FPI.
Close to a mode-hop, at 95\,mA, we clearly see a two-longitudinal-mode operation as seen in fig.\ref{fig:Trait_deux_FP_article} (top).
In the other region, the laser is essentially single-mode, as shown in fig.\ref{fig:Trait_deux_FP_article} (bottom).

\begin{figure}[htp]
\centering
\includegraphics[width=.9\columnwidth]{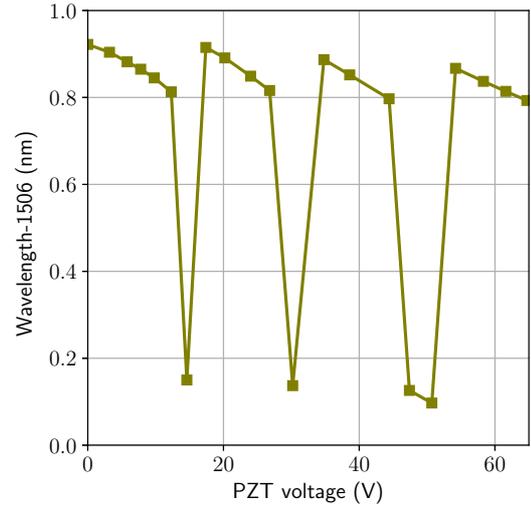}
\caption{Wavelength measurement as a function of the PZT voltage.}
\label{fig:trait_PZT_CW} 
%/home/thierry/Exp/LAC/alcor_TC/20180507_alcor_modulations/
\end{figure}

As a tuning parameter, the PZT voltage can be changed to adjust the cavity length for a fixed injection current (100\,mA in that case).
We obtain in figure~\ref{fig:trait_PZT_CW} the PZT tunablity from the wavelength meter similar to fig.\ref{fig:Trait_series_wavelength_FP}(bottom) for the current.

The wavelength tunability varies from 0.7 to 1.2\,GHz/V (slope of the curve fig.\ref{fig:trait_PZT_CW}).
These typical values are consistent with the previously discussed cavity FSR $\sim$20\,GHz since the PZT displacement is 2.8\,$\mu$m for 75\,V giving the expected value of 1.0\,GHz/V, in good agreement with the experimental values.

%tunability 2.0*2800/1506/75*20 (20GHZ ISL, 2800 deplacement pour 75V)

The CW and longitudinal single-mode operation being validated, we now focus on the laser response to a modulated input for both the injection current and the cavity length (PZT).

\section{Modulation sensitivity responses}

We have intentionally conceived the cavity to perform fast sweeps of the laser frequency.
A short extended-cavity and consequently a large FSR should ensure a good agility when the injection current and PZT are modulated.

\begin{figure}[htp]
\centering
\includegraphics[width=.9\columnwidth]{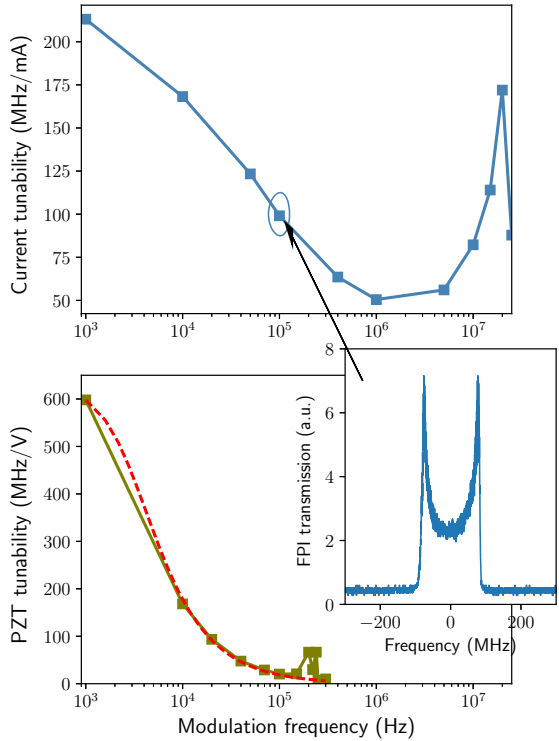}
\caption{Top: Current modulation response in MHz/mA.
Bottom: PZT voltage modulation response in MHz/V.
The dashed red curve represents the expected response limited by the capacitance of the PZT (see text).
Inset: Fabry-Perot interferometer transmission for a 2\,mA current modulation at the frequency 100\,kHz to illustrate the responses measurement method.
The extreme side-bands are separated by $\sim$180\,MHz giving a current sensitivity of 90\,MHz/mA at 100\,kHz (circled point on the current modulation curve).}
\label{fig:trait_modulations} 
%/home/thierry/Exp/LAC/alcor_TC/20180507_alcor_modulations/trait_modulations.py
\end{figure}

The diode injection current is modulated directly though a bias-T by-passing the current controller (500\,$\Omega$ resistor and 3.3\,$\mu$F capacitor, corresponding to a 100\,Hz low-frequency cut-off).
The DC-bias current is then imposed by the diode current controller (100\,mA in this case).
We typically apply a few volts amplitude at the AC input corresponding to few mA of modulation through the 500\,$\Omega$ resistor.
The modulation frequency is varied to characterize the laser response (transfer function).
To evaluate the sensitivity, we use the previously mentioned scanning FPI.
We show in fig.\ref{fig:trait_modulations} (inset) the FPI transmission when the diode current is modulated by 2\,mA at the frequency of 100\,kHz.
By pointing the extreme side-bands, we measure a sensitivity of  90\,MHz/mA at this frequency.

%\begin{figure}[htp]
%\centering
%\includegraphics[width=1.0\textwidth]{trait_current_mod_figure.eps}
%\caption{Fabry-Perot interferometer transmission for a 2mA current modulation at the frequency 100kHz. The extreme side-bands are separated by $\sim$180MHz giving a current sensitivity of 90 MHz/mA at 100kHz.}
%\label{fig:trait_current_mod_figure} 
%\end{figure}

The measurement is reproduced for a wide range of modulation frequencies from 1\,kHz to 25\,MHz to obtain the laser current response (see fig.\ref{fig:trait_modulations}, top).

We observe an unexpected increase of the sensitivity at 20\,MHz modulation.
This resonance cannot be attributed to the expected physical diode laser operation, as for example, relaxation oscillation which usually occurs at few GHz \cite{suhara2004semiconductor}. We suspect instead a purely electronic resonance between internal elements of the laser diode current controller and the bias-T.

The sensitivity at low frequency is in good agreement with the CW operation of the laser as measured in fig.\ref{fig:Trait_series_wavelength_FP} (bottom).
We indeed observe a sensitivity in the range 150-300\,MHz/mA, for the single-mode regions for which the slopes (wavelength as a function of the current in fig.\ref{fig:Trait_series_wavelength_FP}, bottom) vary in this range depending on the driving current.
At higher modulation frequencies, the sensitivity decreases to 50\,MHz/mA (at 10\,MHz).
This value is sufficiently large for fast sweeps of the frequency or a rapid feedback when frequency locking is necessary.

A comparable measurement is performed by acting on the PZT (see fig.\ref{fig:trait_modulations}, bottom).
The low frequency value of the sensitivity (0.6\,GHz/V at 1\,kHz) is in satisfying agreement with CW measurement depicted in fig.\ref{fig:trait_PZT_CW} (from 0.7\,GHz/V to 1.2 GHz/V depending on the PZT bias voltage).
The decrease of the PZT response can be well modeled (dashed red curve in fig.\ref{fig:trait_modulations}, bottom) by considering the capacitance of the actuators.
The capacitor forms a low-pass filter with the input impedance of the modulation signal generator (50\,$\Omega$).
If we leave the capacitance as a free parameter for the data fitting by a low-pass filter frequency response, we obtain a capacitance of 1080\,nF in good agreement with the commercial specifications of the PZT (950\,nF $\pm$ 15\%).
Small resonances are observable around 250\,kHz, quite close from the specified resonant frequency of the unloaded PZT (315\,kHz).
The small mass of the GMRF glued on the ceramics slightly pulls the resonance towards lower frequencies as expected but the impact is quite moderate on the PZT performance.
As a conclusion, the PZT response is essentially limited by the internal capacitance of the actuator.
A well-designed and low-impedance piezo-driver could certainly improve the PZT sensitivity at higher frequencies \cite{pzt_driver}.

As a conclusion, the compactness of the extended cavity allows fast current sweeps (90\,MHz/mA at 100\,kHz, the modulation bandwidth). Additionally, the cavity length tuning (with a PZT) is greatly facilitated by the light mass of the GMRF.
To complete our analysis of the laser, we now evaluate the frequency noise and the long-term stability in typical working conditions.

\section{Noise analysis}

We have chosen a compact setup focused on the frequency agility which requires a short cavity.
Compactness usually ensures a good mechanical stability and a lower sensitivity to thermal drifts.
The tight integration of the optical elements is not the primary goal of our proof-of-principle demonstration but a perspective. Nonetheless, it should be mentioned that the price to pay for a short cavity is the frequency noise.
Following the modified Schawlow-Townes formula \cite{Henry} as a fundamental bound, the quantum-limited linewidth should decrease as the inverse of the length squared, all other parameters being equal (losses and power for example).
We address this question by completing a basic noise analysis.

The frequency noise can be evaluated by performing the commonly-used self-homodyne or heterodyne detection techniques \cite{okoshi1980novel, gallion1984quantum, mercer}.
In our case, we focus on the white-noise component of the laser by using a short delay for the homodyne detection to precisely evaluate the quantum-limited linewidth \cite{ludvigsen1994new, short_homodyne}.
Such a measurement (short delay) doesn't give any information about the 1/f-type low-frequency noise, this latter having a major contribution to the laser linewidth \cite{Kikuchi85,Kikuchi89,exter}. Our prototype is not optimized to reduce the 1/f technical noises, instead we focus on the cavity size (few mm) that should strongly impact the white-noise contribution (high-frequency). This allows to evaluate the impact of the extended cavity as compared to the (non-extended) laser diode.
Additionally, the homodyne detection with a short delay can be easily implemented with standard fibered components allowing perfectly contrasted homodyne beats. There is no need of an acousto-optic modulator in that case.

\begin{figure}[htp]
\centering
\includegraphics[width=.9\columnwidth]{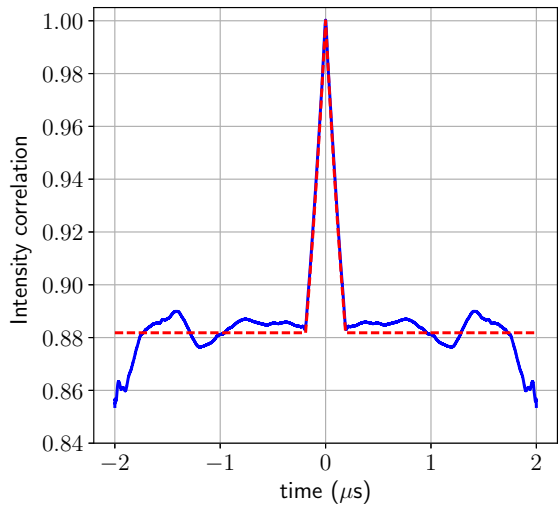}
\caption{Intensity correlation function from self-homodyne detection with short delay (in blue).
The correlation function is well-predicted for short delays by a white-noise model (in dashed red, see text for details). The triangular shape of the curve is not due to a lack of temporal resolution (0.2ns per point) but to the physical shape of the correlation function (see eq.\ref{GE2}).}
\label{fig:Trait_series_homodyne_alcor}
%/home/thierry/Exp/LAC/alcor_TC/20180307_alcor_noise/
\end{figure}

As a reference unbalanced interferometer, we use a fibered Michelson interferometer with a long-delay spool ($\sim$20\,m) on one arm \cite{kefelian2009ultralow} and a short arm.
The measured delay from the spectral fringe spacing is $\tau_0$=190\,ns.
The transmitted intensity is recorded during 2\,$\mu$s and its correlation function is calculated numerically.
As explained in \cite{short_homodyne}, the self-homodyne measurement depends on the path length phase of the interferometer.
To retrieve the intensity correlation function, the phase dependent component should be removed by active phase modulation and subsequent averaging.
In our case, the phase averaging is done passively by repeatedly recording the intensity every few seconds during which the interferometer phase fluctuates from shot to shot. The interferometer doesn't contain any active component at the end.
After 100 averaging, we retrieve the correlation function as plotted in fig.\ref{fig:Trait_series_homodyne_alcor}.

The short delay of the homodyne detection allows to estimate the white-noise contribution to the laser linewidth.
Assuming a purely white-noise $S_0$ for the frequency noise spectrum and neglecting the 1/f-type low-frequency contribution, the intensity correlation function reads as \cite{short_homodyne}:

\begin{equation}
  \mathcal{G}_E^{(2)}\!\left(\tau\right)=\displaystyle  \frac{1}{3}
  \left\{\begin{array}{@{}l@{\quad}l}
   2+e^{-S_0 |\tau|}   & \mbox{if $ |\tau|<\tau_0$} \\[\jot]
   2+e^{-S_0 \tau_0}  & \mbox{if $ |\tau|\ge \tau_0$}
    \end{array}\right.
\label{GE2}
\end{equation}

Eq.(\ref{GE2}) can be used to the fit the measured correlation function in fig.\ref{fig:Trait_series_homodyne_alcor}.
The fit (red dashed line in fig.\ref{fig:Trait_series_homodyne_alcor}) gives a white-noise (Lorentzian) contribution to the laser linewidth of  $S_0=2\pi \times 366$\,kHz.
%Again, this value does not give any information about the linewidth over longer timescales, but at least this shows how beneficial it is to extend the cavity of the diode (1.6\,mm optical length for the diode as estimated in section \ref{cw}) to 7.5\,mm (extended-cavity).
%The expected reduction of the Lorentzian contribution would correspond to a factor 22 (all other parameters being equal) leading to $S_0=2\pi \times 366$\,kHz for the extended-cavity as compared to $\sim$10\,MHz for a typical (non-extended) laser diode with the same power \cite{Kikuchi89}.
This value is smaller than the typical ridge laser diode white-noise limit \cite{Kikuchi89} and shows how beneficial it is to extend the cavity of the diode \cite{exter}.
Again, this value does not give any information about the linewidth over longer timescales (1/f-noise). As the quantum-limited linewidth scales like $1/L_c^2$ ($L_c$ being the cavity length), by extending the cavity from $L_c=1.6$\,mm (ridge diode) to $L_c=7.5$\,mm (extended cavity) we expect a net reduction of the Lorentzian contribution by a factor $7.5^2/1.6^2\simeq 22$ as compared to a ridge laser diode. Thus we estimated that for similar power and same facet reflectivity, the non-extended laser diode would exhibit a $\sim 2\pi \times 8$\,MHz Lorentzian contribution which falls in the range of the typical linewidth for ridge diode laser  as reported in the literature \cite{Kikuchi89}.

%We conclude by recording the long-term drift of the laser wavelength.
%This latter will be essentially limited by the thermal fluctuation of the cavity length.
%We insert a thermoelectric coolers (TEC) under the laser mount (brass colored in fig.\ref{fig:design}) and use our combined current and TEC controller to stabilize the temperature.
%The laser is enclosed in a box to roughly minimize the ambient pressure change.
%In theses typical conditions, we record the wavelength with a WS7-30 wavelength meter during 24 hours (see fig.\ref{fig:Trait_longterm_alcor_WS7_figure}).

%\begin{figure}[htp]
%\centering
%%\includegraphics[width=.8\textwidth]{Trait_longterm_alcor_WS7_figure.eps}
%\includegraphics[width=.7\textwidth]{Trait_longterm_alcor_WS7_figure_cropped.pdf}
%\caption{Long term drifts of the laser frequency over 24 hours.}
%\label{fig:Trait_longterm_alcor_WS7_figure} 
%\end{figure}

As an additional frequency characterisation, we also record the long-term drift of the laser wavelength. We insert a thermoelectric coolers (TEC) under the laser mount (brass colored in fig.\ref{fig:scheme_ensemble},right) and use our combined current and TEC controller to stabilize the temperature. The laser is enclosed in a box to roughly minimize the ambient pressure change. In theses typical conditions, we record the wavelength with a WS7-30 wavelength meter during 24 hours. The maximum drift is $\sim$2\,GHz/h and fluctuate during the day probably  limited by the thermal fluctuation of the cavity length.  The maximum excursion is 13\,GHz. This latter can be significantly improved but is a good starting point in the perspective of a better integration of the laser.

\section{Conclusion}
We have verified the single-mode operation at 1.5\,$\mu$m of an ECDL using a GMRF as external mirror.
As compared to a volume grating as a feedback filter, the GMRF is a planar optical component whose potential in terms of production and integration has still to be investigated. The initial gain in terms of mass and volume comes from the reduction of the substrate thickness from few mm for a volume grating to few hundreds of $\mu$m for planar optics (500 $\mu$m for standard wafer, but much thinner ones are achievable).

As an illustration, we have verified that the modulation of the cavity length by actuating the GMRF position with a PZT is essentially limited by the performance of PZT itself and not the small added mass of the filter.

We explore the regime of short extended cavities (7.5\,mm optical length corresponding to an FSR of $\sim$ 20\,GHz). The short length of the cavity is essentially allowed by the reduced footprint of the GMRF. As we demonstrate, this is a particularly interesting trade-off to obtain a moderate level of white-noise contribution to the laser linewidth (366\,kHz) while maintaining a good modulation sensitivity as confirmed by our measurements when the current (90\,MHz/mA at 100\,kHz, the modulation bandwidth) and the cavity length (through a PZT) are modulated. A longer cavity would reduce the laser noise (following the modified Schawlow-Townes formula if all other parameters stay constant) at the price of lower modulation sensitivities (for both current and PZT). It should be also noted that a straightforward way to reduce the white-noise laser linewidth is to use a higher power semiconductor diode. So the measured white-noise contribution in our proof-of-principle demonstration comes from a technical limitation and not a fundamental one.

Guided mode resonant filters have been investigated in a broad range of situations with many applications in the optical domain \cite{quaranta2018recent}.
When inserted in lasers as intracavity frequency selective mirror, an interesting perspective is the active tunability largely benefiting for the low footprint and the planar design of the guiding structure, especially when electro-optically active materials as substrates or deposited layers are considered \cite{Ichikawa:05, shu2013electro}.
The compactness of the filter scales the future integration of the ECDL and can certainly facilitate the development of low-noise narrow-band and rapidly tunable devices.  

\section*{Acknowledgments}
Authors thankfully acknowledge the LAAS clean room team for technical support and technological expertise provided within the French RENATECH framework. The Laboratoire Aim\'e Cotton have received funding from the Investissements d'Avenir du LabEx PALM ATERSIIQ and OptoRF-Er (ANR-10-LABX-0039-PALM).

\section*{References}

\bibliographystyle{unsrt}
\bibliography{alcor2_proofs}

\end{document}